\documentstyle[preprint,aps,prl,epsf]{revtex}
\begin{document}
\def\bea{\begin{eqnarray}}
\def\eea{\end{eqnarray}}
\def\a{\alpha}
\def\d{\delta}
\def\p{\partial} 
\def\nn{\nonumber}
\def\r{\rho}
\def\rv{\bar{r}}
\def\la{\langle}
\def\ra{\rangle}
\def\e{\epsilon}
\def\o{\omega}
\def\n{\eta}
\def\g{\gamma}
\def\break#1{\pagebreak \vspace*{#1}}
\def\f{\frac}
\draft
\title{Inequivalence of Statistical Ensembles \\in\\ 
Single Molecule Measurements}
\author{Supurna Sinha$^{}$\cite{SUP} and Joseph Samuel$^{}$\cite{SAM}}
\address{Harish Chandra Research Institute,\\
Chhatnag Road, Jhunsi, Allahabad 211 019, India\\
and\\
Raman Research Institute, Bangalore 560 080, India\\}
\maketitle
\widetext
\begin{abstract}
We study the role of fluctuations in 
single molecule experimental measurements of force-extension ($f-\zeta$)
curves. We use the Worm Like Chain (WLC) model to bring out the 
connection between the Helmholtz ensemble characterized by the 
Free Energy ($F(\zeta)$) and the Gibbs ensemble characterized by 
the Free Energy ($G(f)$). 
We consider the rigid rod limit of the 
WLC model as an instructive special case to bring out the issue of 
ensemble inequivalence. 
We point out the need for taking into account the free energy of
transition when one goes from one ensemble to another. 
We also comment on the ``phase transition'' noticed in an 
isometric setup for semiflexible 
polymers and propose a realization of its thermodynamic limit.  
We present general arguments which rule out non-monotonic force-extension
curves in some ensembles
and note that these do {\it not} apply to the isometric ensemble. 

\end{abstract}
\narrowtext
\textwidth=12cm
{\bf Introduction:}

In the past, experiments on polymers were confined to studying their
bulk properties, which involved probing 
large numbers of molecules\cite{bulk}. 
The results of these experiments could be analyzed by using 
the traditional tools of thermodynamics. 
In recent years, however, researchers have been successful in 
micromanipulating {\it single}
biological molecules such as DNA, proteins and RNA to probe 
their elastic
properties \cite{bust}. 
Such studies serve a twofold role. 
On one hand, they shed light on 
mechanical properties of semiflexible polymers,
which are of clear biological importance in processes such as 
gene regulation and transcription \cite{Strick,twist,writhe}.  
On the other hand, they provide physicists with a concrete testing
ground for understanding some of the fundamental ideas of 
statistical mechanics.
In statistical mechanics, an isometric setup would be described by  
the 
Helmholtz free energy, whereas an isotensional setup would be
described by the Gibbs free energy\cite{landau}. In the thermodynamic 
limit 
these 
two descriptions agree, but semiflexible polymers (those with contour 
lengths comparable to their persistence lengths) are {\it not} at the 
thermodynamic limit. Experimentally, both isometric and isotensional 
ensembles are realizable. 
Typically the polymer molecule is suspended (in a suitable medium) between 
a translation stage
and a force sensor.
The force sensor could be realized by using an Atomic 
Force
Microscope (AFM) cantilever or by optical or magnetic forces. 
As noted by Kreuzer and Payne \cite{kreuz},
an isometric setup can be realized using a stiff trap and an isotensional 
setup by using a soft trap. 
In a more sophisticated
version, an electronic feedback circuit is used to control the 
force (or the extension) and one measures the fluctuations in the 
extension
(or the force) \cite{keller}.
Here we will focus on the role of 
fluctuations in single molecule experiments.
In order to correctly interpret such experiments one needs to 
understand the effect of fluctuations on the measured quantities. 
For instance, it turns out, that an experiment in which the 
ends of a polymer molecule are fixed (isometric) and the 
force fluctuates yields a different
result from one in which the force between the ends is held fixed 
(isotensional) and 
the end-to-end distance fluctuates\cite{keller,neumann}. 
This difference can be traced to 
large fluctuations about the mean value of the force or the extension,
depending on the experimental setup. These fluctuations
vanish only in the thermodynamic limit of very long polymers. 

Here we use the Worm Like Chain (WLC) model \cite{kratky,bend,Kleinert} to 
study the inequivalence of ensembles due to 
finite size effects.
The WLC model 
has been very successful in achieving quantitative agreement with  
experimentally measured force-extension curves\cite{bust,Marko}. 
The paper is organized as follows. In Sec. $II$ we discuss the Helmholtz
and Gibbs ensembles. In Sec. $III$ we consider the rigid rod limit which
forcefully brings out the main issues dealt with in this paper. 
In Sec. $IV$ we draw attention to the importance of taking into 
account the free energy of transition in going from one ensemble to 
another. In Sec. $V$ we discuss the thermodynamic
limit of a ``phase transition'' recently seen in semiflexible polymers. 
Finally, we end the paper with a discussion in Sec. $VI$. 
  
{\bf II Helmholtz and Gibbs Ensembles:}

Consider an idealized experiment in which one end of a molecule is 
held fixed at $(x_0,y_0,0)$ and the other end is attached to a dielectric
bead confined to a harmonic optical trap described by the potential
\begin{equation}
V(x,y,z) = A\frac{[(x-x_0)^2+(y-y_0)^2]}{2} + C\frac{(z-z_0)^2}{2}.
\label{trap}
\end{equation}
with $(x_0,y_0,z_0)$ defining the center of the trap. 
Consider $A$ to be small 
so that the bead is free to move in the plane $z=z_0$.
For a polymer of contour length $L$ and persistence length $L_P$
it is convenient to introduce the following dimensionless variables:
$\zeta =\frac{z}{L}$, $\zeta_0 =\frac{z_0}{L}$, $\beta =\frac{L}{L_P}$ and 
$f=\frac{FL_P}{k_BT}$ where $F$ is an applied stretching force and
$k_BT$ is the thermal energy at a temperature $T$.
Consider $P(\zeta)d{\zeta}$, the number of 
configurations (counted with 
Boltzmann weight)\cite{bend} for a polymer of length $L$ 
starting from the origin and ending anywhere on the $x-y$ plane
in an interval $d\zeta$ of $\zeta$. 
The free energy defined by ${\cal F}(\zeta)= -\frac{1}{\beta}ln P(\zeta)$ 
is the 
Helmholtz
free energy.  
The partition function\cite{partfoot} for the combined system consisting 
of the 
polymer molecule
{\it and} the trap is given by:
\begin{equation}
Z(\zeta_0,\beta)=\sqrt{\frac{\tilde C}{2 
\pi}}\int_{-\infty}^{+\infty}{d\zeta e^{-\beta 
{\cal F}(\zeta)}e^{-\tilde{C}\frac{(\zeta 
-\zeta_0)^2}{2}}}
\label{part}
\end{equation}
where $\tilde{C} =C L^2$.
By tuning the longitudinal stiffness $\tilde{C}$ we can realize the two 
limiting
cases. 
${\it Helmholtz:}$
In the limit of a stiff trap ($\tilde{C} \rightarrow \infty$), the 
Gaussian factor 
pertaining to the trap approaches a delta function and one gets
\begin{equation}
Z(\zeta,\beta)= e^{-\beta {\cal F}(\zeta)}
\label{stifftrap}
\end{equation} 
Here we have switched notation to write $\zeta$ in place of $\zeta_0$.
Thus a stiff trap realizes the Helmholtz 
ensemble by constraining fluctuations in the $\zeta$ coordinate.
To extend the molecule from $\zeta$ to $\zeta+d\zeta$ one needs
to apply a force $<f>=\frac{\partial F}{\partial \zeta}$
in order to compensate for the change of entropy. Plotting
$<f>$ versus $\zeta$ we find the $(<f>,\zeta)$ force-extension relation.

${\it Gibbs:}$ 
In the opposite limit of a soft trap ($\tilde{C} \rightarrow 0$ and 
$\zeta_0 
\rightarrow \infty$ such that $\tilde{C}\zeta_0 =\beta f$ 
remains finite),
one gets\cite{partfoot}
\begin{equation}
{\tilde Z}(f,\beta)=\int_{-\infty}^{+\infty}{d\zeta e^{-\beta 
F(\zeta)}e^{\beta 
f\zeta}}
\label{softrap}
\end{equation}
Thus a soft trap permits fluctuations in the $\zeta$ coordinate but 
constrains
the force fluctuations\cite{kreuz} and thus realizes the 
Gibbs ensemble.
${\tilde Z}(f)$ is the generating function for the $\zeta$ distribution.
Defining the Gibbs free energy $G(f)={-\beta}ln{\tilde{Z}}(f)$ we
can work out the mean extension $<\zeta>=-\frac{\partial G}{\partial f}$
and the $(<\zeta>,f)$ force-extension relation.

Notice that ${\tilde Z}(f)$ is the Laplace transform 
of 
$Z(\zeta)$.    
In the thermodynamic limit of long polymers ($\beta 
\rightarrow \infty$) the Laplace transform integral Eq.(\ref{softrap}) 
is dominated by 
the saddle point value and therefore  
${\cal F}(\zeta)$ and $G(f)$ are related by a Legendre transform:
\begin{equation}
{\cal F}(\zeta) = G(f) +f\zeta.
\label{legendre}
\end{equation}
For finite $\beta$ i.e. for a polymer of finite extent, 
the saddle point approximation 
no longer holds true and fluctuations about the saddle point value
of the free energy become important.  
Thus one finds that ${\cal F}(\zeta)$ and $G(f)$ are {\it not} 
Legendre transforms of each other.
We notice that this difference between the Legendre transform [Eq. 
(\ref{legendre})] and 
the Laplace transform [Eq. (\ref{softrap})] is the mathematical origin 
of the finite size fluctuation effects described in this paper. 
These fluctuations are of thermal origin and 
can ultimately be traced to 
collisions of the polymer molecule
with the molecules of the suspending medium. 
In this section we have recovered the results of \cite{kreuz}.
We have also gone beyond \cite{kreuz} and 
traced the origin of the 
difference between the two 
ensembles to the difference between the 
Legendre and Laplace transforms.

{\bf III Rigid Rod Limit: An Instructive Extreme Case}

We noticed in the last section that because of fluctuation effects
the Helmholtz and Gibbs free energies are not related by a 
Legendre Transform. 
An important consequence of this is that the $(F,<z>)$ relation 
is different from the $(<F>, z)$ relation. In other words, the 
force-extension curves plotted in the two ensembles are {\it distinct}
due to finite size fluctuation effects.  
Fluctuations dominate at finite $\beta$ 
and disappear in the thermodynamic limit ($\beta \rightarrow \infty$) of 
flexible polymers. 
We bring out the ensemble dependence 
of the force-extension 
relations explicitly and most dramatically 
in the limiting case of a very stiff polymer ($\beta \rightarrow 0$).

In this extremely rigid limit\cite{frey}, the end of the polymer is uniformly 
distributed over the sphere of directions. In the Helmholtz ensemble we  
thus have $P(z)d{z}= \frac{d z}{2L}$ for $-L<z<L$ and 
$P(z) =0$ otherwise. 
Since the free energy is constant in the range $-L<z<L$ and diverges 
otherwise, we find that the average force $<F>$ vanishes for 
$|z|<L$ and diverges for 
$|z| =L$ (See Fig. $1 a$). In the Gibbs ensemble we find by standard 
manipulations that 
\begin{equation}
<z>=[LCoth{FL}-\frac{1}{F}]
\label{zfrel}
\end{equation}
which differs from the $<F>-z$ relation determined above in 
the Helmholtz ensemble (See Fig. $1 b$). 
The theoretical analysis of the ensemble dependence of the force-extension 
relation based on the rigid rod limit is a new result of this paper.

Thus an experimenter making force-extension measurements on, 
for instance, Actin filaments\cite{loic}, would find that a measurement
in which the force is controlled and the end-to-end distance is measured
leads to a different force-extension curve 
from a measurement in which the end-to-end separation is controlled and 
the force is measured. A theorist interpreting the curves also needs 
to keep in mind whether the curves are obtained in a constant-force
setup or a constant-extension setup since a proper interpretation of
the curves requires a knowledge of the ensemble used in the 
measurement.   

\vspace{.6 cm}
\vbox{
\epsfxsize=13.0cm
\epsfysize=6.0cm
\epsffile{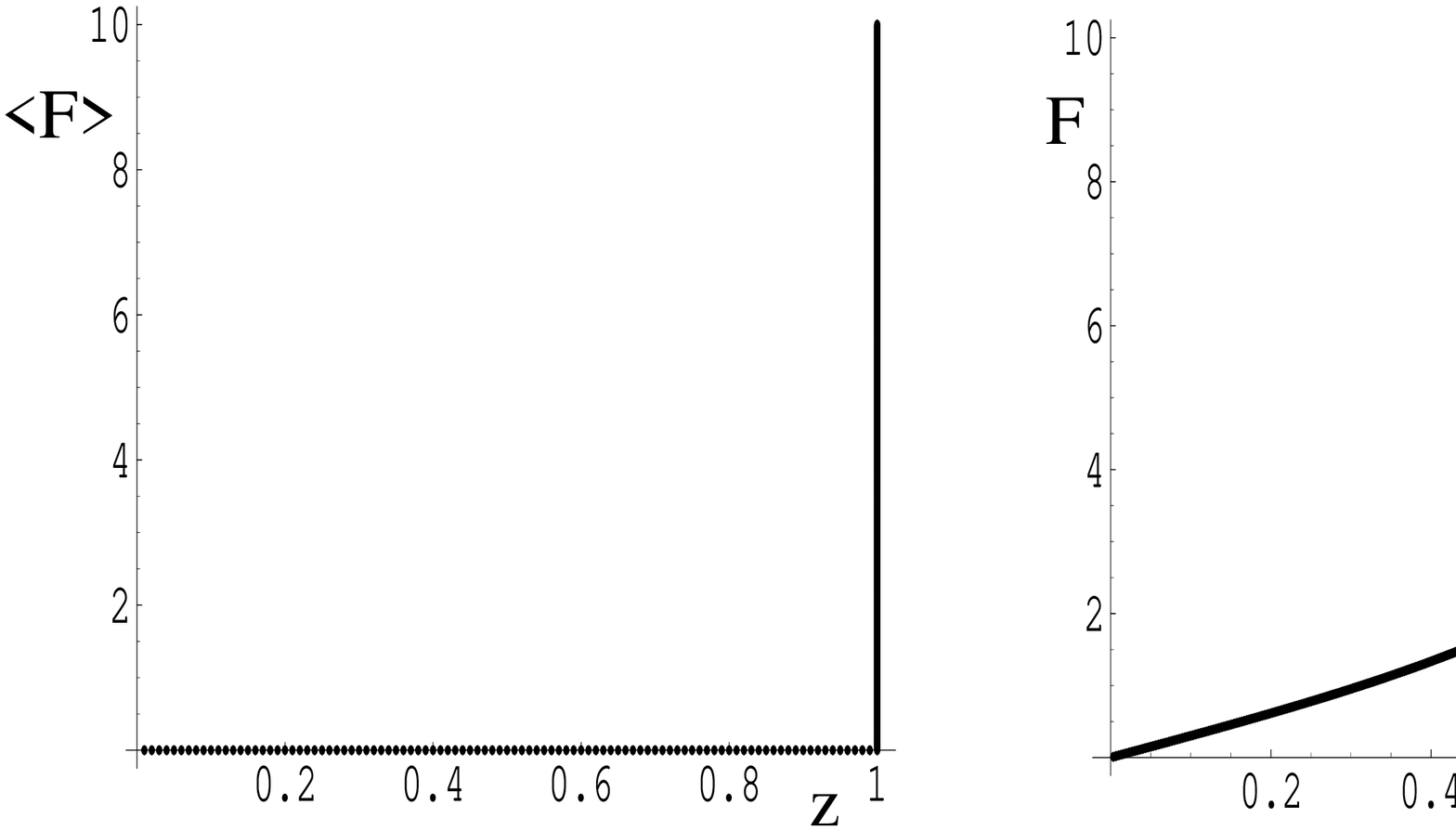}
\begin{figure}
\caption{Force-Extension Curve in the Helmholtz (Fig. 1a) and Gibbs 
(Fig. 1b)
ensembles for $\beta=0.$ We have set $L=1$. }
\label{contrast}
\end{figure}}

{\bf IV Inequivalence Of Ensembles and the Free Energy Of Transition}

In an isotensional setup one controls the force and one 
measures the mean extension and plots
it as a function of force.
In an isometric setup the roles of extension and force are interchanged.
In both setups zero force corresponds to zero extension ($\zeta=0$) 
and a large force corresponds to maximal extension (i.e. $\zeta = 1$).
Imagine going from zero force to a large force via the 
isotensional setup and returning from maximal extension to 
zero extension via the isometric setup.
Since the ``equation of state'' depends on the
chosen ensemble, in general there will be two distinct curves describing
the extension in one ensemble followed by contraction in the other 
\cite{foot}.
In such a situation there could be a net area enclosed in the
Force-Extension plane. 
This poses a puzzle because it appears that a cyclic process can 
extract work from the system. This puzzle is easily resolved. 
In completing the cycle and returning to the initial
state one is in fact changing ensembles twice at the two end points.
These correspond to {\it finite} free energy changes which need to be
taken into account. 
 
Let $(\zeta_1,f_1)$ and $(\zeta_2,f_2)$ be two points which lie on 
both isometric and isotensional force-extension curves. In our example
$(\zeta_1,f_1)=(0,0)$ and $(\zeta_2,f_2)=(1,\infty)$.
Let us suppose that we go from $(\zeta_1,f_1)$ to $(\zeta_2,f_2)$
in the isotensional ensemble and return via the isometric ensemble.
We find that 
\begin{equation}
G(f_2)-G(f_1) =\int_{f_1}^{f_2}{\frac{\partial G}{\partial 
f}df}
=-\int_{f_1}^{f_2}{<\zeta> df}. 
\end{equation}
Similarly,
\begin{equation}
F(\zeta_2)-F(\zeta_1) =\int_{\zeta_1}^{\zeta_2}{\frac{\partial 
F}{\partial 
\zeta}d\zeta}
=\int_{\zeta_1}^{\zeta_2}{<f> d\zeta}.
\end{equation}
The area enclosed between the two curves is given by 
$$W=\int_{\zeta_1}^{\zeta_2}{f d<\zeta>} 
-\int_{\zeta_1}^{\zeta_2}{<f> d\zeta}$$
which can be rewritten as 
\begin{eqnarray}
W=\int_{\zeta_1}^{\zeta_2}{d(f <\zeta>)}
-\int_{f_1}^{f_2}{<\zeta> df} - 
\int_{\zeta_1}^{\zeta_2}{<f> d{\zeta}}\\{\nonumber}
=f_2\zeta_2 - f_1\zeta_1 +G(f_2) - G(f_1)-F(\zeta_2) + 
F(\zeta_1)\\{\nonumber}
=[f\zeta+G(f)-F(\zeta)]_1^{2} 
=[\tilde{F}(\zeta)-F(\zeta)]_1^{2}\\{\nonumber}
=\Delta F^{tr}(\zeta)|_1^{2} 
\label{freetr}
\end{eqnarray}

where we define $\Delta F^{tr}(\zeta)$ as the free energy of transition. 
$\Delta F^{tr}(\zeta)$ is the difference between 
$\tilde{F}(\zeta)=f\zeta+G(f)$,
the Legendre transform of $G(f)$ and the Helmholtz 
free energy $F(\zeta)$. Since these are not equal (except in the limit
of long polymers) the free energy of transition between ensembles must
be considered in order that the total free energy change in a cyclic
process vanishes. 
   
In order to understand this issue more explicitly we consider 
corrections to the saddle point approximation which is valid 
in the long polymer limit.
Let us expand $\phi({\zeta}) = {\cal F}(\zeta)-f\zeta$, the argument of 
the exponential appearing 
on the 
right hand side of Eq. (\ref{softrap})
around the saddle point value $\zeta = \zeta_{*}$ (which dominates the 
integral in 
the long polymer limit) and retain terms upto second order in 
the fluctuations about the saddle point value: 
$$\phi({\zeta}) = \phi({\zeta_{*}}) 
+\frac{1}{2}{\phi}^{''}(\zeta)|_{\zeta=\zeta_{*}}(\zeta 
-\zeta_{*})^2.$$
If we plug in this expansion in Eq. (\ref{softrap}) and identify ${\tilde 
Z}(f,\beta)$ with 
$e^{-\beta 
G(f)}$
we arrive at the following equation:
\begin{equation}
G(f)=[{\cal F}(\zeta_{*})-f\zeta_{*}]+\frac{1}{2\beta}
ln{{\cal F}^{''}(\zeta_{*})}
\label{saddle}
\end{equation}
The free energy due to fluctuations around the saddle point value is
$\frac{1}{2\beta}ln{{\cal F}^{''}(\zeta_{*})}$. 
Notice that in the long polymer limit of $\beta \rightarrow \infty$,
this term vanishes. For finite $\beta$, this nonzero contribution 
to the free energy accounts for the transition between the constant
extension ensemble and the constant force ensemble.
In going from a soft trap to a stiff trap work is done 
on the bead by the trap. Similarly in going from a stiff trap to a soft 
trap work is extracted from the bead by the trap. The net work done is the 
difference between the work done at the two ends
of the force-extension curves in switching ensembles.
This net work exactly cancels out the nonzero 
area enclosed in the force-extension plane. 

{\bf V ``First Order Phase Transition'' and the Thermodynamic Limit in 
Semiflexible Polymers}

In Sec. $II$ we considered the bead to be in a potential which was 
soft in the $x$ and $y$ directions. Let us now consider what happens
when the trap is stiff in all three directions
($A$ as well as $C$ in Eq. ({\ref{trap}}) are large) and the vector 
position 
of the bead is constrained to be at $(x_0,y_0,z_0)$.
Let $Q({\vec r})$ be the number of polymer configurations 
which start at the origin and end in the volume element $d{\vec r}$ 
centered at ${\vec r}$ \cite{bend,dabhi}. 
$Q({\vec r})$ is related to $P(z)$ via the equation 
\begin{equation}
P(z)=\int{d\vec{r} Q(\vec{r})\delta(r_3 -z)} 
\label{pofz}
\end{equation}
which in words, means that $P(z)$ is obtained by 
integrating $Q({\vec r})$ over a plane of constant $z$\cite{bend}.
The distribution $Q({\vec r})$ was studied in \cite{dabhi} 
where it was noticed that 
in an intermediate range (around $3.8$) of $\beta$
the free energy 
$A({\vec r})= {\frac{-1}{\beta}}log{Q(\vec r)}$ had multiple minima.
For a fixed contour length as one varies $\beta$ by tuning the persistence
length $L_P$ one finds a competition between flexible and rigid phases of 
the polymer for intermediate values of $\beta$. Thus the polymer 
undergoes a flexible to rigid ``first order phase transition''
via a two peaked profile of $Q({\vec r})$. 
This leads to a curious 
force-extension relation. 
As one pulls the bead, the restoring force at first increases, 
then decreases to zero and then goes 
negative and becomes a destabilizing force. 
The 
molecule is unstable and goes to a new extended state. 

The words ``first order phase transition'' above were in quotes as a
finite system does not show phase transitions. If one takes the 
thermodynamic limit by taking the length of the polymer to 
infinity ($\beta \rightarrow \infty$) one loses the multiple minima
structure which is present only in a small range of $\beta$ around
$3.8$. Is it possible for this phase transition to survive the 
thermodynamic limit? As we will see below, this is indeed possible 
provided one takes the replica thermodynamic limit. 
We take $N$ replicas of the molecule with fixed $\beta$ and let $N$ 
tend to $\infty$. Consider $N$ identical polymers with $\beta= 
\frac{L}{L_P}$ fixed, their two ends anchored to flat surfaces 
$S_1$ and $S_2$  (See Fig. $2$). One could realize the above
arrangement by using 
$(i)$ two planar arrays of optical traps or
$(ii)$ by introducing suitably synthesized supramolecular lamellar
structures. 
The anchoring is such that the tangent vectors to the molecule 
at the fixed ends are free to swivel. If one applies a 
force $F$ to pull $S_1$ and $S_2$ 
apart the $N$ molecules are also stretched. We consider the molecules
to be well separated so that they can be regarded as independent.
One could now look
at the mean force $F$ needed to maintain the separation $r$. It is easily
seen
that the force is proportional to $N$ and also the mean square
fluctuation
$<(\Delta F)^2>$ in the force is proportional to $N$. 
This is because the mean force and its variance are 
respectively the first and second derivatives of the free energy,
which being an extensive quantity is proportional to $N$. 
It follows that as
$N$ goes to infinity, the fluctuations ($\Delta F/F$) in $F$ die out as
$1/\sqrt{N}$. We can now regard the mean force $F$ as a control parameter
(i.e. consider a constant $F$ ensemble)
and observe that if we tune the applied force $F$, there is a
discontinuous change in the
separation $r$ between the two sets of optical traps 
signalling a first order phase transition with the inter-trap separation
$r$ as the order parameter. Thus, 
in the replica way of taking the thermodynamic limit 
the double humped
form of the distribution function  $Q({\vec r})$ results in a true first 
order phase transition. 

\vbox{
\epsfxsize=6.0cm
\epsfysize=6.0cm
\epsffile{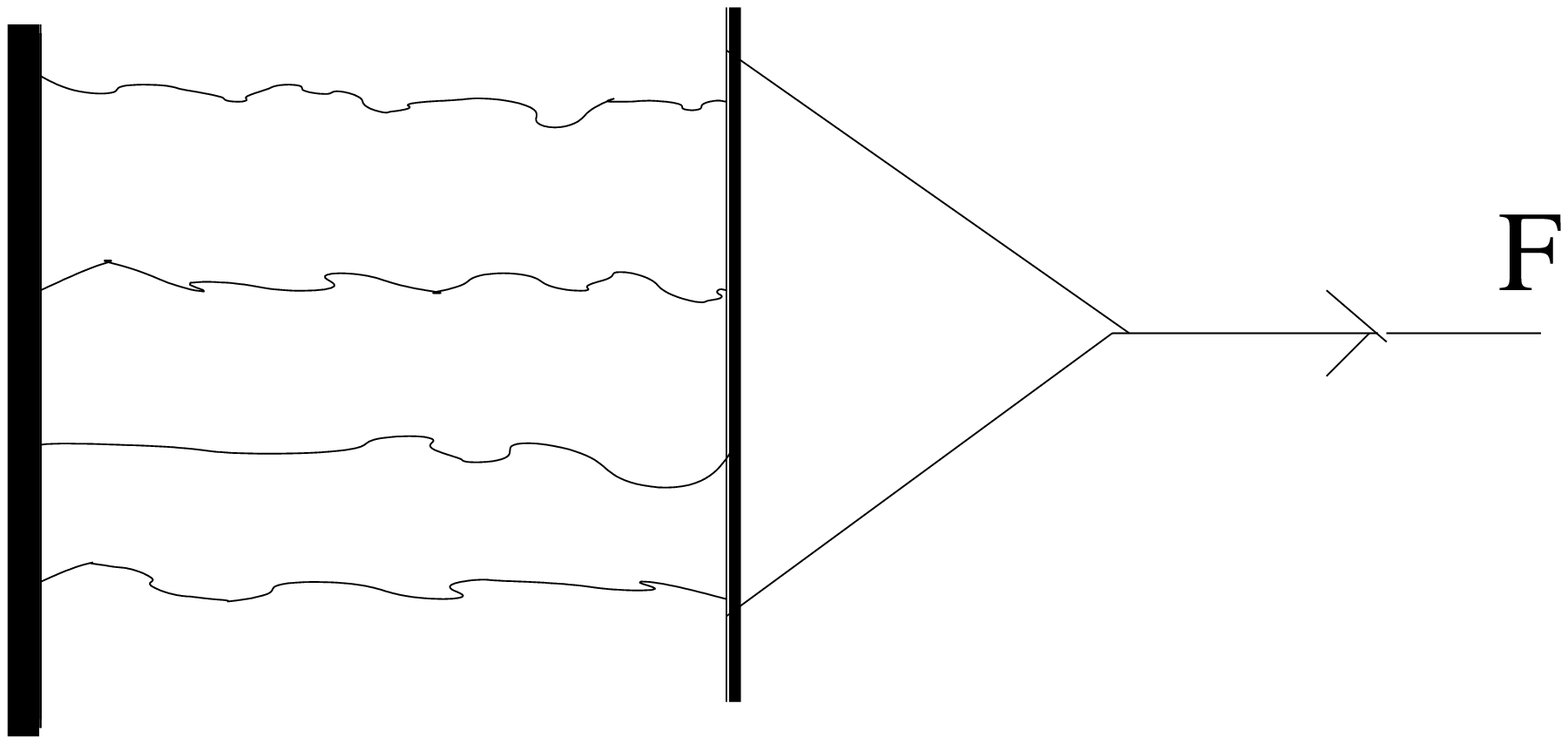}
\begin{figure}
\caption{Schematic Experimental Design for Replica Thermodynamic Limit,
shown above for 
$N=4.$}
\label{replica}
\end{figure}}

{\bf VI Discussion:}

In this paper we point out the importance of considering the free energy 
of transition in going between the Helmholtz and the Gibbs ensembles
in the context of single molecule force-extension measurements. 
We also study two distinct ways of taking the thermodynamic limit
$(i)$ by letting the length of the polymer tend to infinity
(i. e. $\beta \rightarrow \infty$) and 
$(ii)$ by considering replicas. In particular we notice that 
the flexible to rigid transition mentioned in Sec. $V$ for a 
single semiflexible polymer survives the replica thermodynamic limit. 
In contrast, this feature disappears in the 
usual thermodynamic limit of $\beta \rightarrow \infty$. 

The non-monotonic behavior of $Q(\vec r)$ is an intriguing feature of 
molecular elasticity. 
It was noticed in recent simulations\cite{dabhi} and subsequently in 
a semi-analytical treatment\cite{bend}. It was also commented on in a 
recent paper\cite{Kleinert2}.
This double humped form of $Q({\vec r})$
has gained renewed interest in the context of cyclization 
of polymers and its significance to gene regulation \cite{gautam,CELL}.

We note that this remarkable qualitatively
distinctive feature is special to the isometric ensemble (fixed vector
end-to-end separation) and is not shared by other ensembles. 
In other ensembles one can argue generally that
such non-monotonic behavior cannot occur. For example, the conjugate
distribution 
${\tilde P}({\vec F})=\int Q({\vec r})e^{{\vec F}.{\vec 
r}}d{\vec r}$ 
is monotonic. This follows from noticing that the $3\times 3$ matrix  
\begin{equation}
\frac{\partial{ln {\tilde P}}}{\partial{F_i}\partial{F_j}}
=<(r_i -<r_i>)(r_j -<r_j>)> 
\label{nonmon}
\end{equation}
is positive definite. Such 
arguments do not apply to $Q({\vec r})$ since there is no analogous
formula to Eq. (\ref{softrap}) in the conjugate distribution. Indeed,
if there were, a double humped form could not appear in $Q({\vec r})$,
for one could express the second derivative of $Q({\vec r})$ as the 
variance of the force.
Can $P(z)$ 
show non-monotonic behavior? The answer is no,
for it has been shown in \cite{bend} that  
$${\frac{-2}{z}}{\frac{d P}{d z}} = Q(z)$$
Since $Q(\vec r)$ is a probability density and therefore non-negative,
it follows that $\frac{d P}{d z} \leq 0$ for $z>0$, thus 
ruling out multiple peaks in $P(z)$. 
This argument which rules out multiple peaks in $P(z)$ is a new 
observation. 
Note that a $P(z)$ measurement differs from $Q(\vec r)$ only in the 
transverse stiffness $A$ of the trap. By tuning $A$ we can permit 
fluctuations in the transverse direction and therefore 
destroy the 
phase transition present in the stiff $A$ limit.
One would expect to see a critical stiffness $A=A_c$ below which the 
phase transition is destroyed. Alternately, one could tune the mean 
force and expect to see the phase transition vanishing below a critical 
mean force $F=F_c$ for a fixed value of $\beta$ in the intermediate
range of $\beta$.  
We emphasize that the non-monotonic 
features of $Q({\vec r})$ in the semiflexible range $\beta \approx 3.8$
are predictions of the WLC model which can be tested against experiments.
A single molecule with its ends confined in optical traps is expected
to show this flexible to rigid transition. The effect can be dramatic
however, if a large number of molecules co-operatively show such a 
transition. One could attach the ends of 
a collection of semiflexible
polymers to supramolecular 
layers\cite{drozdov} and detect the 
flexible to rigid
transition signalled by a change in the inter-layer 
separation via a suitable probe. 
It may be possible to exploit this dramatic transition from flexible to 
rigid behavior in technological applications.   

If one considers the replica thermodynamic limit of a semiflexible polymer
one sees that force-extension curves continue to remain distinct in 
the Helmholtz and the Gibbs ensembles. 
So while interpreting a force-extension curve obtained for a collection 
of semiflexible polymers suspended between two arrays of traps,
one needs to know if the curve is obtained using
the soft trap setup or a stiff trap setup.
However, for $\beta \rightarrow 
\infty$, which is the usual thermodynamic limit, the two ensembles 
give rise to the {\it same} force-extension curve.
This observation is consistent with 
the fact that inequivalence of ensembles can 
survive 
at the thermodynamic limit for systems with long range 
interactions\cite{kreuz,keller,Lynd}. 
In the context of semiflexible polymers,
the persistence length $L_P$ plays the role of 
the range of interactions.
There has been some work\cite{baro} on the thermodynamics
of particle systems in the presence of external macroscopic 
fields in classical and quantal contexts. In these papers
the authors have dealt with the macroscopic limit of the 
definition of pressure which is analogous to the thermodynamic 
limit of the definition of force in our work. 
In particular the authors of Ref. \cite{baro} discuss the 
connection between the values of the
pressure defined by two different thermodynamic limit procedures:
in the first, the system is confined successively in a
sequence of boxes which grows to fill up the whole space.  In the 
second,
the system is in an external potential 
similar to the present context.
In the case of a
semiflexible polymer the force corresponding to a given extension is 
the same in the Gibbs and Helmholtz ensembles only in the thermodynamic
limit of $\beta \rightarrow \infty$. This is analogous to the second
procedure of taking the thermodynamic limit 
given in \cite{baro}, where they recover the thermodynamic notion 
of pressure from an underlying microscopic definition when they 
let the scale factor go to infinity in the macroscopic limit. 
The results of this paper are therefore, consistent with the general 
treatment in \cite{baro}. 

The fluctuation effects mentioned here also 
have some biological significance.
In particular, the process of gene regulation involves interaction
between parts of a DNA molecule which are about 
less than one persistence length
apart ($\approx 34 nm$)\cite{CELL}. Over such short segments of the 
DNA fluctuation effects would be significant.
The replica thermodynamic
limit would also play a role in the concrete biological 
context of a network of  
actin filaments forming the cytoskeletal structure.

{\it Acknowledgements:}
We thank Dipanjan Bhattacharya for drawing attention to Ref. 
\cite{keller}
and Abhishek Dhar, Deepak Dhar, Erwin Frey and Richard Neumann for 
discussions.
We thank one of the referees for drawing attention to Ref.\cite{baro}.
We thank Harish-Chandra Research Institute for their hospitality and 
the wonderful working atmosphere in which this paper was written.


\begin{references}
\bibitem{partfoot} We drop additive constants in the free energies or
equivalently, multiplicative constants in the partition functions.
\bibitem{foot}In fact we have illustrated this point explicitly in the 
context of the rigid rod limit in Sec. $III$.
\end{references}
\end{document}